\documentclass[12pt,american]{article} 
\usepackage[T1]{fontenc} 
\usepackage[latin9]{inputenc} 
\usepackage{babel} 
\usepackage[noblocks, affil-sl]{authblk} 
\usepackage{graphicx}
\usepackage{amssymb}
\usepackage{amsmath}

\usepackage{booktabs} 

\usepackage{todonotes}

\usepackage[textwidth=6.5in, top=1in]{geometry}
\usepackage{setspace}

\usepackage[figurename=Fig., labelsep=period]{caption}

\begin{document}

\title{Digital representation and quantification of discrete dislocation networks}

\author[a]{Andreas E. Robertson}
\author[a,b]{Surya R. Kalidindi \footnote{Email: surya.kalidindi@me.gatech.edu} \footnote{Corresponding Author}}

\affil[a]{George W. Woodruff School of Mechanical Engineering, Georgia Institute of Technology, Atlanta, GA 30332 USA}
\affil[b]{School of Computational Science and Engineering, Georgia Institute of Technology, Atlanta, GA 30332 USA}

\maketitle

\begin{abstract}
Dislocation networks and their evolution are known to control the mechanical properties of metal samples. However, the lack of computationally efficient and statistically rigorous descriptors for such defect systems has hindered the development and adoption of rational protocols for the optimal design of these material systems.  This study presents a framework for the rigorous statistical quantification and low dimensional representation of dislocation networks using the formalism of 2-point spatial correlations (also called 2-point statistics) along with Principle Component Analysis (PCA). The usefulness of this basic framework for comparing and observing dislocation networks is exemplified and discussed with suitable examples. 
\end{abstract}

\newpage
\onehalfspacing

\section{Introduction}

Materials exhibit rich features over a hierarchy of length scales, spanning from the atomistic to the macroscale \cite{mcdowellplasticity, elliot, cpfem}, which collectively control their bulk properties. For example, a material's bulk plastic properties are strongly influenced by mesoscale features (e.g., crystallographic texture, grain size and shape distributions) \cite{cpfem, mcdowellplasticity, druckermultiscale, mcdowellviscoplast} as well as crystal defects at the lower length scales (e.g., dislocations) \cite{mcdowellplasticity, capDDDsurvey, azabpo, greer}. In order to attain the improved combinations of properties demanded by advanced technologies, it is necessary to engineer these desired salient features in the context of the multiscale material structure. Rigorous approaches to address these materials design challenges are critically dependent on access to high-fidelity materials knowledge expressed as salient process-structure-property (PSP) linkages. This materials knowledge base is traditionally accessed via computationally expensive physics-based simulations (e.g., crystal plasticity finite element models \cite{cpfem}, phase field simulations \cite{phasefield, PCevolution}) or by experimental observations. Because successful materials design efforts typically require repeated access to the underlying knowledge base, the high cost of these methods makes it impractical to deploy them directly in current design protocols. At the mesoscale, it has been shown that the challenges described above can be addressed using the data-centric Materials Knowledge System (MKS) framework \cite{kalidbook1, fullwoodsurvey}. This framework provides systematic workflows to mine desired high-value materials knowledge, contained in all available experimental and simulation data \cite{pymks}, into low-computational cost surrogate models. Recent advances in physics-based simulation tools (e.g., three-dimensional (3D) Discrete Dislocation Dynamics (DDD) \cite{capDDDsurvey, azabpo}) and characterization techniques (e.g., Transmission Electron Microscopy (TEM) \cite{tem, kachertem, originaltem}) for studying dislocation networks have demonstrated the capacity for generating and aggregating similarly large datasets. However, comparable frameworks to MKS to address the material design needs at the dislocation length scales do not yet exist. As a result, new research avenues have opened for the development and application of suitable data science approaches for studying and extracting high value knowledge from dislocation datasets. 

Quantitative microstructural statistics form the basis for materials data science methods \cite{fullwoodsurvey,kalidbook1,torquato}. As a result, the identification and adoption of salient statistics is an essential first step in the rigorous study of dislocation networks. Such descriptors should be efficient to calculate, easy to visualize and compare, and capture the spatial arrangement of important dislocation features \cite{fullwoodsurvey}. This final requirement extends from the close relationship between the spatial arrangement of dislocations and their bulk behavior and/or properties \cite{hirthlothe, hullbacon}. In the general study of dislocations, several potential candidate statistics are commonly used. The bulk dislocation density is most popular because it is easy to calculate and compare. Additionally, it can be used to predict bulk properties, such as the Critical Resolved Shear Strength (CRSS), with modest success \cite{taylorrule}. Other, less commonly used measures, include fractal analysis \cite{zaiser} and spatial correlations \cite{AzabDeng, wang, kacher, anderson, groma1, groma2}. Generally speaking, the bulk dislocation density is insufficiently descriptive because it does not describe the spatial arrangement of the network. On the other hand, fractal statistics are overly specialized: they are mainly meant to represent cell forming dislocation networks. 

In contrast, several studies have illustrated the promise of spatial correlations in describing effectively a network's salient attributes \cite{AzabDeng, wang, kacher, anderson}. However, these previous investigations also highlighted some of the current deficiencies. Primarily, the complex functional form and high dimensional nature of the correlations largely precludes meaningful visualization and analysis without significant simplifications (e.g., Deng and El-Azab \cite{AzabDeng}). They remain prohibitively high dimensional for tasks such as classification of large data-sets and extraction of high fidelity PSP linkages. Secondly, these earlier efforts have not yet proposed an efficient computational framework for extracting network statistics from discrete dislocation networks (such as those generated in DDD simulations). 

The primary goal of this paper is to propose and demonstrate a computational framework for the extraction of low-dimensional salient spatial statistics from discrete dislocation networks. The central tasks in this development include the establishment of a generalized mathematical framework for the spatial correlations of dislocation networks, the construction of a systematic and efficient scheme for their computation, and the extraction of salient, low dimensional statistics from the high dimensional spatial correlations. This paper presents a framework addressing these gaps, and demonstrates its merits through simple examples. 

\section{Background}
\subsection{Spatial Correlations} \label{spatial_correlations}

N-point spatial correlations (also referred to as n-point statistics) \cite{kalidbook1,torquato,fullwoodsurvey,orig2PS,pymks} offer the most versatile and comprehensive framework for the rigorous statistical quantification of the complex hierarchical internal structures encountered in most advanced materials. Indeed, these concepts have been employed successfully to study material structures at the mesoscale in a broad of range of applications from classification of microstructures \cite{taxonomy, eutectic, molecdynamics} to the extraction of high-fidelity low-computational cost PSP linkages \cite{PCevolution, nonlinear2ps, fatigue2ps, polycrystal2ps} calibrated to physics-based simulations or experimental observations. They are also beginning to be employed on atomistic datasets \cite{matthew}.

In the mesoscale theories, n-point statistics were first encountered in Kroner's statistical treatment of the effective stiffness tensor of a two-phase composite microstructure \cite{kroner}. This treatment proposed a quantitative relationship between the n-point statistics of the local elastic stiffness and the effective elastic stiffness. This approach was subsequently generalized to a broad range of the microstructure's effective properties \cite{kalidbook1, torquato}. These statistical theories formulate this connection by interpreting the specific spatial arrangement of microstructural features as a sampled instance from a stationary, spatially resolved, ergodic random process  \cite{azabStochFibProc2006, AzabDeng, stochmicro, kalidbook1}. Within this statistical interpretation, a microstructure with multiple local states (such as a two-phase composite) is generated from multiple co-dependent random processes. A natural conclusion of this interpretation is that the effective properties are dependent on the random process, not the observed microstructure instantiation. Because such a random process is completely characterized by its n-point statistics, the desired connection is established naturally \cite{randproc, stochmicro}. 

In practice, a microstructure's set of 2-point statistics are often a sufficient descriptor \cite{PCevolution, polycrystal2ps, azabStochFibProc2006, groma2}. For a simple illustration of their definition, let us consider a spatially resolved $H$-phase microstructure random process, $M^h(\boldsymbol{x})$. Here, $h$ serves as an integer index from 1 to $H$, indexing the random process for each discrete material local state. For two arbitrary states, $\beta$ and $\gamma$, the corresponding 2-point statistics are defined by 
\begin{equation}
    {}_Mf^{\beta \gamma}(\boldsymbol{\tau}) = E[M^\beta(\boldsymbol{x}+\boldsymbol{\tau})M^\gamma(\boldsymbol{x})]
    \label{eq:corrgendef}
\end{equation}
\noindent Because the processes are assumed to be stationary, the 2-point statistics are spatially resolved functions of the difference vector between two points, $\boldsymbol{\tau}$. As a result, the 2-point statistics are defined by the expectation of the processes at points spatially separated by $\boldsymbol{\tau}$. Because microstructure random processes are assumed to be ergodic, the expectation in Eq. (1) is equal to the spatial average of the product $\left\langle m^\beta(\boldsymbol{x}+\boldsymbol{\tau})m^\gamma(\boldsymbol{x})\right\rangle$ over a sampled instance, $m^h(\boldsymbol{x})$. Mathematically, this is expressed as \cite{randproc}
\begin{equation}
    {}_Mf^{\beta \gamma}(\boldsymbol{\tau}) = \left\langle m^\beta(\boldsymbol{x}+\boldsymbol{\tau}) m^\gamma(\boldsymbol{x})\right\rangle = \frac{1}{X} \int_X m^\beta(\boldsymbol{x}+\boldsymbol{\tau})m^\gamma(\boldsymbol{x})d\boldsymbol{x}
    \label{eq:corrdef}
\end{equation}
\noindent At the mesoscale, Kalidindi \textit{et al.} refer to $m^h(\boldsymbol{x})$ as the microstructure function \cite{stochmicro, kalidbook1}. It mathematically represents the sampled instance from the microstructure random process, $M^h(\boldsymbol{x})$.  In Eq. (\ref{eq:corrdef}), $X$ denotes the volume of the representative volume element (RVE).

In order to simplify the computation defined in Eq. (2), Kalidindi \textit{et al.} define a discrete microstructure function, $m^h_s$, by voxelizing the spatial domain \cite{kalidbook1, fullwoodsurvey}. The discrete microstructure function takes an average value for each voxel, indexed by $s$. The corresponding  discrete 2-point statistics are defined by  
\begin{equation}
    {}_Mf_t^{\beta \gamma} = \frac{1}{S} \sum_{s=1}^S m_{s+t}^\beta m_s^\gamma
    \label{eq:corrdiscrete}
\end{equation}
\noindent In this discrete form, the functional dependence on $\boldsymbol{\tau}$ is replaced by a discrete index of the voxel separation, $t$ \cite{kalidbook1}. Instead of averaging over the domain volume, $X$, the discrete spatial statistics are averaged by the total number of feasible samples, $S$. For periodic microstructures, $S$ is the same as the number of voxels. These concepts can be extended to higher-order spatial correlations (i.e., beyond 2-point statistics in the framework of n-point statistics). Because of their expected causal relationship with the microstructure's effective properties, n-point statistics serve as natural structural descriptors within the PSP framework \cite{kalidbook1}.

The considerations described above are also applicable at the scale of discrete dislocations. With a suitable selection of local state descriptors, the spatial arrangement within a dislocation network can also be treated as a random process, and can be rigorously quantified using the formalism of n-point statistics \cite{anderson, azabStochFibProc2006, AzabDeng, groma1, groma2}. Furthermore, these measures can be quantitatively correlated with any of the effective (homogenized) properties defined for dislocation networks \cite{sauzay, temporal_stat}. 

\subsection{Spatial Description of Dislocation Networks} \label{spatial_descriptions}
Previous studies have explored different mathematical representations of the dislocation networks for capturing the spatial distributions of both individual dislocations and suitably grouped ensembles of dislocations \cite{groma1, groma2, AzabDeng, azabStochFibProc2006, hochrainer, sedlacek, giessen, valdenaire, arsenlis, kroner_dislocation}. These definitions trade off complexity for accuracy in their representation. For example, the spatially resolved dislocation density, $\rho(\boldsymbol{x})$ provides a simple and easy to understand description (the total dislocation line length per unit volume, defined in the limit of a small volume in the neighborhood of the spatial location, $\boldsymbol{x}$). A more accurate descriptor can be defined by further subdividing the density onto each slip system at the spatial location, $\rho^k(\boldsymbol{x})$. Here, $k$ indexes the slip-system. 

Recently, the following spatially resolved vector density representation has gained prominence \cite{arsenlis, azabStochFibProc2006, AzabXia2015, groma1, sedlacek}:  
\begin{equation}
    \boldsymbol{\rho}^k(\boldsymbol{x}) = \rho^k_{b}(\boldsymbol{x}) \hat{b} + \rho^k_{t}(\boldsymbol{x}) \hat{t} + \rho^k_{n}(\boldsymbol{x}) \hat{n} \ \ k = 0, 1, 2, ..., K
    \label{eq:densityvector}
\end{equation}
\noindent For each slip system, the vector local state of the dislocation network is defined using an orthobasis comprised of the screw (or burger's) direction, $\hat{b}$, the edge direction, $\hat{t}$, and the out-of-plane direction, $\hat{n}$. The corresponding components, $\rho^k_i(\boldsymbol{x})$, reflect the geometrically necessary dislocation densities, GND, along each direction. The GND is defined as the net dislocation length per unit volume \cite{arsenlis}. It has been shown that this descriptor contains sufficient information to calculate stress fields induced by the dislocation network \cite{AzabXia2015, bertin, azabpo} and to formulate closed field equations for the network's evolution \cite{AzabXia2015, azabpo}. Additionally, the spatial correlations of this vector density field have been used to define the source terms controlling the network's evolution \cite{temporal_stat}. This representation does have some limitations because it reflects only the net dislocation content at the spatial location $\boldsymbol{x}$. As an example, considering a small volume in the neighborhood of $\boldsymbol{x}$ containing two dislocations on the same slip system and having equivalent length but opposite vector direction (or "sense"), the GND would be zero. Clearly, the selection of the neighborhood size implied in this definition needs careful consideration. 

\subsection{Digital Representation, Fourier Methods, and Dimensionality Reduction}
As discussed earlier, the growth of data-scientific techniques to study, compare, and extract high-fidelity knowledge from discrete dislocation data is limited by a lack of efficient tools for extracting quantitative statistical descriptors of the observed networks and methods for easily representing them. We leverage several tools from data science, computer graphics, and, digital signal processing to overcome these obstacles. 

\subsubsection{Siddon's Algorithm}
For voxel-based  descriptions of a dislocation network, it becomes necessary to calculate the line length contained within each voxel. The voxelization of straight lines is an area of extensive research in computer graphics. Siddon's algorithm, used extensively in medical imaging, traces the voxels crossed by a 3D straight line and records the length traversed within each voxel \cite{siddons}. Additionally, Siddon's algorithm exhibits computational cost complexity O(N) making it efficient for the voxelization of large datasets, such as those generated by DDD. In this work, the original algorithm is also suitably modified to calculate vector line density without significantly increasing its computational cost.

\subsubsection{Fourier Methods for Calculating 2-Point Statistics} \label{fouriermethods}
Previous research by Kalidindi \textit{et al.} has extensively documented that voxelized representations of material structures facilitate computational efficiency gains throughout the learning of PSP linkages by application of the algorithms and techniques of digital signal processing \cite{kalidbook1}. For collections of large datasets, such efficiency is paramount. A significant benefit of voxelization observed at the mesoscale is that 2-point statistics can be efficiently calculated using the Fast Fourier Transform, Eq. (\ref{eq:fouriercorr}) \cite{kalidbook1}. 
\begin{equation}
    {}_Mf_t^{\beta \gamma} = \frac{1}{S} \sum_{s=1}^S m_{s+t}^\beta m_s^\gamma = \left( \frac{1}{S} \mathbb{F}^{-1} \left[ \mathbb{F} \left[ m^{\beta} \right] \mathbb{F}^* \left[ m^{\gamma} \right] \right] \right)_t
    \label{eq:fouriercorr}
\end{equation}
\noindent $\mathbb{F}[\cdot]$ denotes the Discrete Fourier Transform, $\mathbb{F}^{-1}[\cdot]$ is its inverse transform, and $\mathbb{F}^*[\cdot]$ is its complex-conjugate. As before, the discrete 2-point statistics remain indexed by the voxel separation, $t$. In contrast to the $O(N^2)$ time complexity of implementing Eq. (\ref{eq:corrdiscrete}), the Fast Fourier Transform (FFT) algorithm facilitates the equivalent calculation, via Eq. (\ref{eq:fouriercorr}), in $O(NlnN)$ time. Because of the reduction in computational complexity, usage of the FFT makes the  calculation of spatial statistics affordable for large collections of datasets. 

\subsubsection{Dimensionality Reduction}
Dimensionality reduction techniques from digital signal processing and machine learning allow distillation of high-dimensional data to salient low-dimensional representations. Principle Component Analysis (PCA) \cite{stochmicro, polycrystal2ps, taxonomy, fatigue2ps, molecdynamics} addresses this task by mean-centering the data and applying a rigid rotation to define a data-driven orthonormal basis organized along the directions of decreasing unexplained variance \cite{princeton, DSP}.  Dimensionality reduction is achieved by utilizing only the weights of the high variance basis vectors.  This technique is favored for several reasons. First, because the transformation is simply a rigid rotation, the euclidean distance metric is preserved. The only error in the distance between data points is the information lost due to truncation of the basis. Second, the information loss is easily quantifiable. This loss is captured by the explained variance along each lost axis. Finally, PCA can be efficiently calculated using Single Value Decomposition making it feasible for very large datasets \cite{princeton, DSP}.

\section{Feature Extraction for Dislocation Networks}
Leveraging the theory and tools introduced in the previous sections, we now propose a flexible framework for extracting salient, low dimensional statistics (i.e., compact representations of spatial correlations) for collections of discrete dislocation networks. First, in the context of the discrete microstructure function paradigm, we propose an efficient scheme for calculating 2-point statistics from discrete dislocation data. Next, using PCA, we extract salient (i.e., low dimensional) network statistics from the calculated 2-point statistics. We envision that this framework will nurture the study, comparison, and extraction of high fidelity knowledge from dislocation network data. We assume that the raw data for our computations is available in the format favored by DDD simulations, as a directed set of linear segments within a simulation domain. The different protocols employed in the proposed framework are described next.

\subsection{Voxelization}
\subsubsection{Definition of the Local State}

First, the simulation's volumetric domain is discretized into a uniform voxel grid. The selection of the local state describing the contents of each voxel is a pivotal step in the proposed framework. Primarily, its selection dictates the accuracy with which the discrete dislocation network is represented as a voxel grid and, by extension, the ability of the generated statistics to differentiate dislocation networks. In this work, we combine the ideas described in Sec. \ref{spatial_correlations} and  Sec. \ref{spatial_descriptions} to arrive at a consistent framework for describing dislocation networks. Specifically, it is recognized that quantities such as\footnote{In contrast to past studies, we have preferred the dislocation length ($l$) to the dislocation density ($\rho$). In voxelized representations, these are simply related to each other by the voxel volume, which is a constant in our representation.}${}^{,}$\footnote{Using the square root of the dislocation length as a local state affords a more natural interpretation for the calculated correlations. This is because the correlations have units of length. Specifically, the $t=0$ auto-correlation for this choice can be related to the dislocation density through a constant factor.} $l$, $l^k$, $l^{1/2}$ and $(l^k)^{1/2}$, where $k$ identifies a specific slip system, offer excellent choices for the local state variable. In other words, one can use $l_s^k$ as $m^h_s$ in defining the spatial correlations (see Eq. (\ref{eq:corrdiscrete})) of interest in the dislocation networks.  However, the framework offers additional flexibility. For example, one can use $\boldsymbol{l}^k =\{l^k_{b}, l^k_{t}, l^k_{n} \}$ (see Eq. (\ref{eq:densityvector})) as the local states. This would, of course, improve the descriptive accuracy of the dislocation network (by specifying individually the screw, edge, and climb components of each slip system) at an increased computational cost. For example, for an FCC crystal, the $(l^k)^{1/2}$ description results in 12 local states per voxel (one per slip system). In contrast, the vector description, $\boldsymbol{l}^k$, results in 36 local states. In this work, we will show that spatial arrangement of many dislocation networks can be differentiated using just $(l^k)^{1/2}$.

\subsubsection{Defining a Reference System} \label{reference_system}
Crystal symmetry poses certain challenges in assigning a reference frame to any given dislocation network. For example, there are 24 completely equivalent assignments of the reference frame to a cubic crystal. One way to address this is to replicate and represent each dislocation network in each of its equivalent reference frames. While this strategy does address the problem, it also increases the variance in the overall dataset. The second, more preferred, option is to standardize the assignment of the crystal reference frame such that it provides an automated labeling order for all of the available slip systems based on specified criteria. For example, in the case studies presented in this work, the crystal reference frame is selected such that the slip system with the highest bulk dislocation density is labelled as the $\left(111\right)\left[\bar{1} 1 0\right]$ slip system. Additional criteria may be needed in other case studies. For example, if a dislocation network exhibits the exact same bulk dislocation density on two of its slip systems, one would need additional criterion to assign the crystal reference frame. 

\subsubsection{Voxelization Density}
A careful selection of the voxel density is critical to ensuring a sufficiently accurate representation of the dislocation network. In general, this selection must balance computational feasibility with representational accuracy. In the mesoscale representations \cite{kalidbook1, PCevolution, eutectic}, the voxel density is often dictated by the size of the smallest microscale constituent (i.e., phase regions) in the material structure. However, the one-dimensional (1D) nature of dislocations precludes a simple extension of these prior protocols. Xia and El-Azab \cite{AzabXia2015} have recommended the interaction length of annihilation as a discretization length scale for dislocation based simulations. The annihilation distance varies from around 2 nm for edge dislocations to about 50 nm for screw dislocations \cite{Annihil, AzabXia2015}. In this study, we used a domain size of $(2000b)^3$ ($b$ is the magnitude of the burgers vector) with a voxelation of $151\times151\times151$. This corresponds to a voxel size of about $13b$, falling within the range discussed above.  

The careful selection of voxel size is very important. Because of our interest in the use of PCA for the extraction of the salient features, it becomes necessary to adopt a single voxel size for all dislocation networks within the dataset. This standardization allows for a meaningful comparison of the calculated 2-point statistics of the different networks \cite{ahmet, molecdynamics}.

\subsubsection{Discrete Representations} \label{voxelization_application}
A discretized representation of the dislocation network is obtained using the tools and concepts presented earlier. We will exemplify the steps involved using the vector local state in an FCC crystal. As already noted, this discretized representation involves the specification of 36 local states for each voxel in the dislocation network volume. This task can be addressed in a computationally efficient protocol by leveraging the ray-tracing algorithm described earlier \cite{siddons}. For vector local states, the inner products of the unit directions of each linear dislocation segment in the network with the basis vectors ($\{\hat{b}, \hat{t}, \hat{n}$\}) of all twelve slips systems are first computed. Then, the dislocation line length assigned to each traversed voxel is calculated using Siddon's algorithm \cite{siddons} and multiplied by the already computed direction cosines to obtain the desired dislocation descriptors introduced in Eq. (\ref{eq:densityvector}). Completing this process for all of the line segments in the DDD dataset then results in the desired voxelized description of the dislocation network with 36 local states for each voxel. The other scalar local states described earlier are similarly populated without the usage of the direction cosines. 

\subsubsection{Kernel Smoothing} \label{kernel_smoothing}
Our initial attempts at implementing the protocols described above revealed that the voxelized representations are extremely noisy. Here, the term noise refers to the large changes to the discretized fields caused by small changes in the voxel size. This problem can be addressed by smoothing the discretized fields, where the contribution of each line segment is spread over the voxel it is passing through and its nearest neighbors. This practice has been implemented successfully in many image analyses protocols \cite{imagesmooth, alsmooth} as well as in the quantification of atomistic data-sets \cite{matthew}. For dislocation structures, Cai's nonsingular burger's vector kernel (Eq. (\ref{eq:caikernl})) \cite{cai, bertin} provides a physics-based selection of the smoothing kernel, parameterized by the radial spread, $a$. The kernel is mathematically expressed as
\begin{equation}
    \tilde{w}(x) = \frac{15}{8\pi} \times \left[ \frac{1-m}{a_1^3 \left( r^2/a_1^2 + 1\right)^{7/2}} + \frac{m}{a_2^3 \left( r^2/a_2^2 + 1\right)^{7/2}} \right] 
    \label{eq:caikernl}
\end{equation}
with $a_1=0.9038a$, $a_2=0.5451a$, and $m=0.6575$ \cite{cai}. In the case studies presented in this work, we have decided to set $a=50b$. We found this value sufficient to remove significant noise and stabilize the statistics, while remaining sufficiently below the average length of individual dislocations to maintain their linear character.

Application of the kernel in Eq. (\ref{eq:caikernl}) to the voxelized structure can be done efficiently via the FFT. For the end goal of calculating 2-point statistics, this smoothing is achieved at minimal additional cost by extension of Eq. (\ref{eq:fouriercorr}) as 
\begin{equation}
    {}_Mf_t^{\beta \gamma} = \left( \frac{1}{S} \mathbb{F}^{-1} \left[\mathbb{F} \left[ \tilde{w} \right] \mathbb{F} \left[ m^{\beta} \right] \mathbb{F}^* \left[ m^{\gamma} \right] \mathbb{F}^* \left[ \tilde{w} \right] \right] \right)_t
    \label{eq:filteredfourier}
\end{equation}
\subsection{2-Point Statistics}
\subsubsection{Calculation}
The protocols described above have transformed each dislocation network into $H$ discrete spatial fields. Treating each field as a local state, the network's auto- and cross-correlations can be efficiently calculated using  Eq. (\ref{eq:filteredfourier}). For a system with $H$ states, there exist $H^2$ correlations.

\subsubsection{Correlation Independence}
Although $H^2$ correlations can be calculated,  Niezgoda \textit{et al.} \cite{neizgoda} have shown that this complete set exhibits many interdependencies. Eq (\ref{eq:interrelation}) defines the derived interrelationship between the Discrete Fourier Transforms (DFT) of the 2-point statistics. 
\begin{equation}
    {}_MF_t^{\beta\gamma} = \frac{({}_MF_t^{\alpha\beta})^*{}_MF_t^{\alpha\gamma}}{{}_MF_t^{\alpha\alpha}}
    \label{eq:interrelation}
\end{equation}
\noindent From this relationship, it is clear that any correlation can be calculated from a single reference state's auto-correlation and that state's cross-correlations with all other states. As a result, only $H$ independent correlations exist. The reference slip-system selected in Sec. \ref{reference_system} should also be used as the as the reference local state (in the presented case studies, the highest dislocation density slip-system is used). We note that, in practice, the nonlinearity of these interdependencies often mandates inclusion of more than the minimum correlations in successful efforts aimed at learning the underlying knowledge in the datasets. Specifically, inclusion of all the autocorrelations can significantly improve performance. 

\subsection{Dimensionality Reduction}
Even after removal of the interdependencies, the dimensionality of the complete set of 2-point statistics is intractably large. For a voxelization of $151\times151\times151$ and $H=36$ set of correlations, the number of 2-point statistics is over 120 million; this results in a unwieldy set of features representing each dislocation network. PCA allows for the dimensionality reduction needed before any further analysis (e.g., classifying dislocation networks, building PSP models). Consider a data set containing N dislocation networks, each having H local states and S voxels. For each dislocation network, we can define a feature vector of size $1\times(SH)$ containing the network's vectorized 2-point statistics (see Eq. (\ref{eq:datavector})). Let the elements of this vector be denoted ${}_ig_j^{\beta\gamma}$ , which indicates a scaled 2-point statistic of dislocation network $i$ for voxel separation vector index $j$ and local states $\beta$ and $\gamma$. It is crucial to follow a consistent method for unrolling the 3D 2-point statistics throughout a dataset. This consistency allows the data vector generated for each structure to be compared. 
\begin{equation}
    x_i = \left[ {}_ig_1^{11},..., {}_ig_S^{11},..., {}_ig_1^{\beta\gamma},..., {}_ig_S^{\beta\gamma},... \right]; \ {}_ig_j^{\beta\gamma} = \frac{{}_if_j^{\beta\gamma}-\mu_j^{\beta\gamma}}{\sigma_{\beta\gamma}}
    \label{eq:datavector}
\end{equation}
\begin{equation}
    \mu_j^{\beta\gamma} = \frac{1}{N} \sum_{n=1}^N {}_nf_j^{\beta\gamma}
    \label{eq:mean}
\end{equation}
\begin{equation}
    \sigma_{\beta\gamma} = \left( \frac{1}{S} \sum_{s=1}^{S} \frac{1}{N} \sum_{n=1}^N ({}_nf_s^{\beta\gamma}-\mu_s^{\beta\gamma})^2 \right) ^ {1/2}
    \label{eq:variance}
\end{equation}
\noindent The vectorized feature list in Eq. (\ref{eq:datavector}) is mean-centered (see Eq. (\ref{eq:mean})) and scaled (see Eq. (\ref{eq:variance})) in preparation for PCA. Because PCA identifies and aligns the data basis with the orthogonal directions of maximum variance, the application of PCA is highly sensitive to differences in magnitudes between the concatenated features. Smaller magnitudes will generally result in lower variance. For example, the magnitudes of auto-correlations are generally higher than those of the cross-correlations. Therefore the application of PCA directly on ${}_if_j^{\beta\gamma}$ is very likely to emphasize more the features with higher magnitudes. This poses a challenge, because one would ideally desire to afford equal importance to each set of spatial correlations. The normalization shown in Eq. (\ref{eq:variance}) ensures that each set of spatial correlations corresponding to a selected pair of local states ($\beta,\gamma$) gets the same attention in the PCA. 

Finally, to perform PCA and extract the desired low dimensional statistics, the N individual network vectors in a dataset are stacked to form a data matrix:  
\begin{equation}
    \mathbf{X} = \begin{bmatrix} {}_1g_1^{11} & \cdots & {}_1g_S^{11} & \cdots \\ \vdots &  \vdots & \vdots & \vdots \\ {}_Ng_1^{11} & \cdots & {}_Ng_S^{11} & \cdots \end{bmatrix}
    \label{eq:datamatrix}
\end{equation}
\noindent PCA is most commonly performed by Singular Value Decomposition (SVD) of $\mathbf{X}$ \cite{princeton}. The low dimensional statistics for each data-vector (for each dislocation network) correspond to the weights (or PC scores) of the highest variance principle components. 

\section{Case Studies in Computations of Dislocation Network Statistics}
We present two case studies designed to illustrate the merits of the proposed protocols for computing and comparing different dislocation networks based on their spatial correlations. In data-driven feature engineering approaches such as the one presented here, the results will depend on the dataset itself. It is therefore important to aggregate a diverse collection of dislocation networks exhibiting suitable levels of interclass and intraclass variabilities. Collecting such datasets from experiments or DDD simulations is quite challenging. Therefore, for the present study, it was decided to use digitally created artificial dislocation networks exhibiting distinct local and global features, resulting in distinct classes of dislocation networks. These artificially created dislocation networks provide an excellent test bed for exploring and understanding the benefits and limitations of the feature engineering protocols developed in this work. 

\subsection{Generation of Dislocation Networks}

\begin{figure}[h]
    \centering
    \includegraphics[scale=0.45]{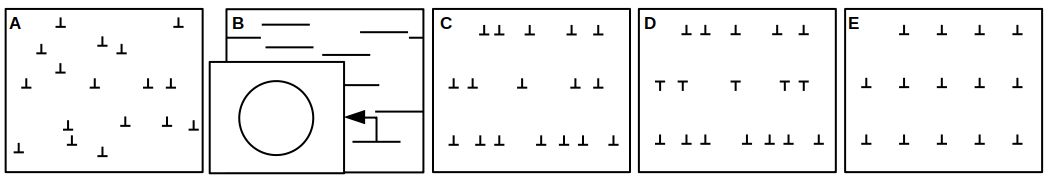}
    \caption{Schematics of the archetypes used in the generation of the different classes of dislocation networks identified in Table \ref{tab:classdescrip}. From left to right: Straight dislocations randomly distributed throughout the volumetric domain (Type A). Dislocations loops randomly distributed throughout the volumetric domain (Type B). Straight dislocations randomly populated on uniformly spaced slip planes (Type C). Straight dislocations randomly populated on uniformly spaced slip planes with alternating orientations on neighboring planes (Type D). Straight dislocations uniformly populated on uniformly spaced slip planes (Type E).}
    \label{fig:classes}
\end{figure}

Fig. \ref{fig:classes} exemplifies the archetypes used to generate the 11 classes of dislocation networks analyzed in this work. The different archetypes are exemplified in 2D sections for simple visualization, although the generated dislocation networks are 3D. In the 2D schematics shown in Fig. \ref{fig:classes}, the slip plane extends perpendicular to the image. The 11 classes of dislocation networks generated for this study are differentiated at two scales: local and global. Locally, networks are differentiated by the preferential shape and orientation of individual dislocations. As depicted in Fig. \ref{fig:classes}, Types A, C, and E are locally identical in that they contain straight, uniformly oriented, dislocation segments with a specified length. The dislocation segments in these archetypes can be of either uniformly screw or edge character (only uniformly edge networks are depicted in Fig. \ref{fig:classes}).  In contrast, Type B is locally differentiated by its circular dislocation loops. Some of the generated networks are globally differentiated by preferential spatial arrangements of dislocations. For example, Types A and B were created with their dislocations randomly arranged throughout the domain. In contrast, the dislocations in Types C, D, and E were preferentially populated on uniformly spaced planes. The alternating sense of individual dislocations on the neighboring planes in Type D differentiates them from the ones in Type C (which contains dislocations of uniform sense). Type E displays an additional level of uniformity in the placement of dislocation segments on each plane. All the dislocation networks were produced in a FCC crystal lattice and periodicity was imposed in the generation. 

The archetypes in Fig. \ref{fig:classes} have been used to generate the 11 different classes of dislocation networks listed in Table \ref{tab:classdescrip}, along with the relevant parameters used in their generation. Each class in this table is labelled as X:Y:N. X refers to the type of dislocations (e.g., edge (E), screw (S), loop (L)). The modifier $15$ following the X label indicates that the dislocations were populated with length $1500b$. All others contained dislocations of length (or circumference) $1000b$. Y refers to either random placement (R) or the number of uniformly spaced planes selected for the placement of the dislocations. Letters d and e have been added to the middle label to indicate the placement strategies identified as Type D and Type E in Fig. \ref{fig:classes}, respectively. N refers to the number of different slip systems on which the dislocations were placed. This label has been forgone for classes with a single slip system. For the present study, we have limited our attention to dislocation networks on only one or two slip systems. This simple strategy for generating networks allows us to systematically explore the effect of populating additional slip systems on the computed spatial statistics. For classes with multiple slip systems, the second system (SS2) was populated with a constant relative position to the first (SS1). In the protocols used in this work, SS1 refers to the slip system with the highest dislocation density, and is made to always correspond to $(111)[\bar{1}10]$ slip system. For each class, multiple instances were generated while maintaining the parameters identified in Table \ref{tab:classdescrip}, but with changes to the seeding of the dislocations.

\begin{table} 
\centering
\begin{tabular}{l p{12mm} c p{20mm} p{70mm}}
\toprule
& \multicolumn{4}{c}{\textbf{Generation Parameters}} \\ 
\cmidrule(l){2-5}
\textbf{Class Label} & Local & Global & Dislocation \newline Length & Additional Notes\\ 
\midrule 
E:R & Edge & Random & $1000b$ & Class A. Populated on SS1 \\ 
E15:R & Edge & Random & $1500b$ & Class A. Populated on SS1 \\ 
S:R & Screw & Random & $1000b$ & Class A. Populated on SS1 \\ 
L:R & Loop & Random & $1000b$ & Class B. Populated on SS1 \\ 
E:11 & Edge & 11 Planes & $1000b$ & Class C. Populated on SS1. \\ 
E:14 & Edge & 14 Planes & $1000b$ & Class C. Populated on SS1. \\ 
E:11d & Edge & 11 Planes & $1000b$ & Class D. Populated on SS1. \\ 
E:11e & Edge & 14 Planes & $1000b$ & Class E. Populated on SS1. \\ 
E:R:2 & Edge & Random & $1000b$ & 60\% populated on SS1 (edge), 40\% populated on SS2 (edge) (\% of total length). Class A on both planes. \\ 
S:R:2 & Screw & Random & $1000b$ & 60\% populated on SS1 (screw), 40\% populated on SS2 (screw) (\% of total length). Class A on both planes. \\
ES:R:2 & Edge \newline Screw & Random & $1000b$ & 60\% populated on SS1 (edge), 40\% populated on SS2 (screw) (\% of total length). Class A on both planes. \\ 
\bottomrule 
\end{tabular}
\caption{Details of the 11 classes of dislocation networks generated for this study.}
\label{tab:classdescrip}
\end{table}

\subsection{2-point statistics} \label{analyze_2ps}

\begin{figure}[h]
    \centering
    \includegraphics[width=129mm]{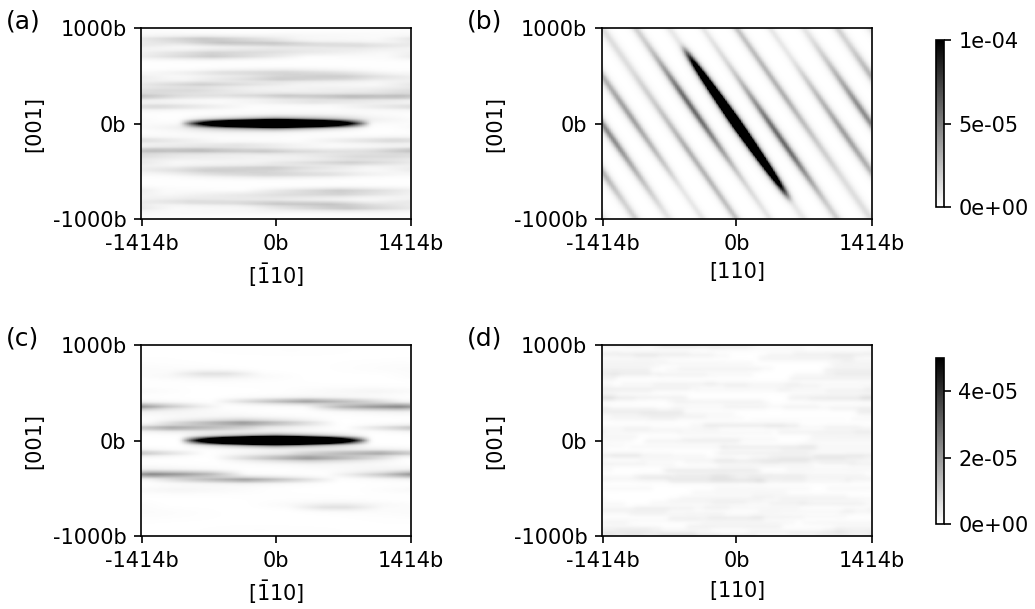}
    \caption{Cross sections from correlation maps. (a) Auto-correlation map on the $(110)$ plane for SS1 dislocations in an example network from S:R class. (b) Auto-correlation map on the $(\bar{1}10)$ plane for SS1 dislocations in an example network from E:11 class. (c) Auto-correlation map on the $(110)$ plane for SS1 dislocations in an example network from S:R:2 class. (d) Cross-correlation map on the $(\bar{1}10)$ plane between SS1 and SS2 dislocations in an example network from S:R:2 class.}
    \label{fig:2ps}
\end{figure}

The auto- or cross-correlations computed for each dislocation network can be visualized as 3D maps as they reflect the value of a statistic for a selected voxel separation vector, $t$ (see Eq. (\ref{eq:corrdiscrete})). Fig. \ref{fig:2ps} presents the correlation maps for three example dislocation networks. For clarity, we present only 2D cross-sections of these 3D maps. Fig. \ref{fig:2ps}(a) displays the $(110)$ cross-section from the auto-correlation map computed for SS1 dislocations for an example network from the S:R class. Fig. \ref{fig:2ps}(b) displays the $(\bar{1}10)$ cross-section from a class E:11 network's SS1 auto-correlation map. Figs. \ref{fig:2ps}(c) and \ref{fig:2ps}(d) display the $(110)$ cross-section from the SS1 auto-correlations and the $(\bar{1}10)$ cross-section from the SS1-SS2 cross-correlations for a class S:R:2 example network. We note that each presented section represents only a small subset of the full set of 3D 2-point statistics computed for each network. These have been specifically selected to illustrate the main features in the computed 2-point statistics. Voxelization was accomplished using $(l^k)^{1/2}$ local states for the two populated slip systems. The domain size for all networks was $(2000b)^3$, and contained dislocations totaling $30000b$ in length. 

The most easily interpreted statistic in the auto-correlation map is the one corresponding to  $t=0$ (see the center of the auto-correlation maps in Fig. \ref{fig:2ps}(a)-(c)). With the use of $(l^k)^{1/2}$ as the local state, the $t=0$ auto-correlation corresponds to the expected dislocation line length in a voxel for the $k^{th}$ slip system. For Fig. \ref{fig:2ps}(a) and Fig. \ref{fig:2ps}(b), the $t=0$ auto-correlation value\footnote{Although the networks' bulk density was maintained at a constant value, slight differences in the $t=0$ auto-correlation value arise from differences in voxelization and the applied kernel smoothing protocols described in Sections \ref{voxelization_application} and \ref{kernel_smoothing}.} was $\sim 4.5\cdot 10^{-4}b$. In contrast, the $t=0$ SS1 auto-correlation for the S:R:2 network was $\sim 2.7\cdot 10^{-4}b$, which is 60\% of the value for the single slip system networks. Since the S:R:2 network exhibits a 60-40 distribution of the total line length between the two slip systems, the observations above indicate that the computed statistics are quite accurate. The relatively small magnitude of the average dislocation length compared to the voxel size of $(13.24b)$ highlights the volumetric sparcity of the dislocation networks used in this study. While the simple networks allow us to interpret their 2-point statistics maps more easily, their volumetric sparcity makes them more sensitive to computational noise.  

The dominant shape in the center of the auto-correlation map (see Fig. \ref{fig:2ps}(a)-(c)) carries information on the average shape and orientation of the local features in the dislocation network. The high correlation region in the center of these maps is likely to come from connected voxels in individual features in the network. For example, Fig. \ref{fig:2ps}(a) depicts the auto-correlations from a class S:R network containing $1000b$ dislocation segments solely directed along the slip direction. Consequently, for any voxel containing a dislocation in this network, we expect to find additional voxels containing dislocations by traversing along the slip direction, with the probability of success dropping as we move far away from the initially specified voxel. This expectation quantitatively manifests as the central elliptical feature oriented along the screw direction in Fig. \ref{fig:2ps}(a). This feature's length, $\sim 2000b$, reflects the fact that the probability of finding a voxel with a dislocation essentially goes to zero when one traverses $1000b$ along the slip direction in this network (in both positive and negative directions along the slip direction). The feature's intensity decreases from a maximum at $t=0$ to a minimum at either tip. The auto-correlation map in Fig. \ref{fig:2ps}(a) also indicates that the probability of finding another voxel with a dislocation decreases sharply for traversals in any other direction. Additionally, the lighter horizontal bands in the auto-correlation map in Fig. \ref{fig:2ps}(a) capture the stacking of the dislocations in the parallel planes. The random arrangement of these secondary bands reflects the fact that the planes for the placement of the dislocations in the S:R network were selected randomly.

Similarly, the auto-correlation map presented in Fig. \ref{fig:2ps}(b) captures the features expected for E:11 networks. In these networks, the dislocation lines are placed along the $[11\bar{2}]$ crystal direction (because these are edge dislocations on the $(111)[\bar{1}10]$ slip system). The central elliptical feature in this figure can be interpreted the same way as before. In contrast to Fig. \ref{fig:2ps}(a), the secondary bands are much stronger and display uniform spacing. These features reflect the fact that, for class E:11 networks, the dislocations were populated on uniformly spaced planes. The distances between these bands (approximately $\sim 289b$) matches the spacing of populated planes in the generation of this network.

The auto-correlation and cross-correlation maps in Fig. \ref{fig:2ps}(c) and \ref{fig:2ps}(d) capture the 2-point statistics for an example dislocation network in the S:R:2 class. As expected, the SS1 auto-correlation map is visually quite similar to the auto-correlation map for the S:R network in Fig. \ref{fig:2ps}(a). This is because the same strategy was used in placing the dislocations on the SS1 slip systems in both these networks. The main difference, as already noted, is that the correlations are roughly scaled down by 60$\%$, reflecting the lower overall density of SS1 dislocations in the S:R:2 network. The good comparison of the features in the auto-correlation maps in \ref{fig:2ps}(a) and \ref{fig:2ps}(c) confirm that these maps mainly reflect the generation parameters and not the seeding of individual dislocations in creating a specific instantiation of the dislocation network. The main difference between the one and two slip system networks is actually seen in the cross-correlation maps. For a network with a single slip system, the cross-correlation map would only show zero values. For the two slip system network, the cross-correlation map presents information on spatial correlations between the dislocations on the two slip systems. For example, Fig. \ref{fig:2ps}(d) displays the expected SS2 neighborhood surrounding a voxel containing an SS1 dislocation for an example S:R:2 network. The map displays faint linear features extending in the $[110]$ crystal direction. Such features indicate that when a voxel containing an SS2 dislocation is observed near a SS1 containing voxel, it is highly probable that more SS2 containing voxels can be found by traversing in the $[110]$ direction. This reflects the fact that, for S:R:2 networks, SS2 dislocations were $(1\bar{1}1)[110]$ screw dislocations. Furthermore, the random arrangement of the SS2 dislocations with respect to the SS1 dislocations is quantitatively captured in the random placement of these linear features. 

We note the fluctuations observed in the present secondary features. These fluctuations are a consequence of under sampling in the spatial averaging used to approximate the network correlations, Eq. (\ref{eq:corrdef}). For example, in Figs. \ref{fig:2ps}(a) and \ref{fig:2ps}(c), the divergence from the nearly uniform long-range correlation we would expect from globally random networks, illustrates that the spatial averaging is still capturing some of the specific seeding of the network instance. In general, this behavior is a consequence of the networks' sparsity and the 1-D nature of the individual dislocations. These characteristics make it difficult to spatially sample a sufficient number of features to accurately calculate the network class's correlation. In the subsequent case studies, these fluctuations will account for the observed intraclass variance. In Sec. \ref{discussion}, we will present several methods for minimizing this noise. 

\subsection{Case Study 1}

This case study is aimed at demonstrating the ability of the proposed statistics to automatically distinguish (i.e., conduct unsupervised classification) generated networks exhibiting differences in both their bulk dislocation density as well as their archetype (see Fig. 1). For this purpose, dislocation networks of classes E:R, S:R, and L:R were generated at three distinct total lengths: $30000b$, $90000b$, and $150000b$, resulting in nine different classes of networks. For each class, 30 instantiations were made using the same generation strategy but changing the seeding of the dislocations in the networks. The dislocation density values were chosen to mimic the densities observed in realistic dislocation networks \cite{capDDDsurvey}. 

\begin{figure}[h]
    \centering
    \includegraphics[width=129mm]{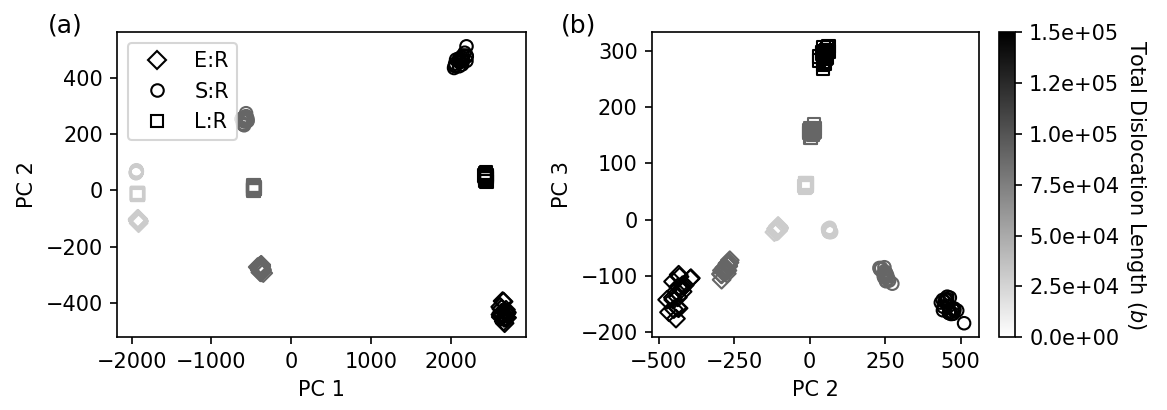}
    \caption{Representation of the generated dislocation networks in the reduced-order PC space. (a) PC1-PC2 subspace, and (b) PC2-PC3 subspace. Each point in these plots represents a dislocation network, with the shape of the data point referring to the dislocation archetype and the gray-scale reflecting the total dislocation length in the network.}
    \label{fig:pcCS1}
\end{figure}

Fig. \ref{fig:pcCS1} presents the 2-point statistics of the generated networks in reduced PC subspaces. Since the networks were not labelled in any manner before performing the PCA, it is satisfying to observe that the proposed protocols are able to successfully separate the dislocation networks into classifiable groupings. It is also clearly seen that the intraclass variance (capturing the differences between multiple networks of a given class) is significantly smaller than the interclass variance (i.e., differences between the networks produced using different generation parameters). Fig. \ref{fig:pcCS1}(a) illustrates that the first PC score largely captures differences in the dislocation line length (corresponding to the $t=0$ auto-correlation). It must also be noted that the first PC score will not exhibit an exact one-to-one exclusive mapping to the network's total dislocation length. Instead, this information is likely distributed throughout all the PC scores in very different ways. For example, Fig. \ref{fig:pcCS1}(b) shows that the distance from the origin in the PC2-PC3 subspace is also likely to correlate well with the overall dislocation density in the network.

In addition to discriminating the networks by their dislocation densities, the protocols used in this work have successfully identified additional differences between the multiple classes of networks. It appears that PC2 is reflecting the differences in the dislocation archetypes (i.e, edge, screw, and loop) used for generating the networks. Specifically, it is satisfying to see that the PC2 scores for L:R are about midway between the PC2 scores of E:R and S:R, thereby recognizing that the loops are indeed made of roughly equal amounts of screw and edge components. It is remarkable that the protocols identified this feature automatically (i.e., in an unsupervised setting). Furthermore, the PC3 scores appear to specifically separate the loop archetype from the archetypes containing straight dislocations. 

The dimensionality reduction offered by the protocols in the example shown here is quite remarkable, from about 3.4 million 2-point statistics to 3. In this specific example, the first three PC scores explained roughly 75\% of the total variance in the data-set. This means that the higher PC scores still contain fairly significant information. However, the interpretation of the features captured by the higher PC scores becomes increasingly difficult, since each PC score still represents a weighted combination of about 1.1 million 2-point statistics for this case study. Also, given the volumetric sparsity of the dislocation networks studied here, some of the higher PC scores simply capture the computational noise, which is a part of the intraclass variance seen in Fig. \ref{fig:pcCS1}. It is emphasized here that the clean separation of the signal and white noise highlights one of the important benefits of PCA. As has been documented extensively in digital signal processing literature \cite{DSP, ekg}, PCA acts as a denoising filter. For the study of dislocation networks, where the sparsity induces significant noise, this property is especially valuable.  

\subsection{Case Study 2}
As the second case study, we will demonstrate the ability of the proposed protocols to discriminate dislocation networks based solely on their local and global generation parameters. In other words, we will apply the protocols to an ensemble of dislocation networks that exhibit the same bulk dislocation density but very different local and/or global dislocation placement characteristics (see Fig. \ref{fig:classes} and Table \ref{tab:classdescrip}). The total dislocation line length was maintained at $30000b$ for all networks generated for this case study. Twenty networks were generated for each of the 11 classes listed in Table \ref{tab:classdescrip}. It is emphasized that all 220 dislocation networks produced for this case study are essentially indistinguishable if dislocation density was used as the sole network quantification metric. Employing the computational protocols presented in this work, the SS1 auto-correlations, the SS2 auto-correlations, and the SS1-SS2 cross-correlations were computed and included in this analyses ($\sim 10$ million 2-point statistics).  

\begin{figure}[h]
    \centering
    \includegraphics[width=174mm]{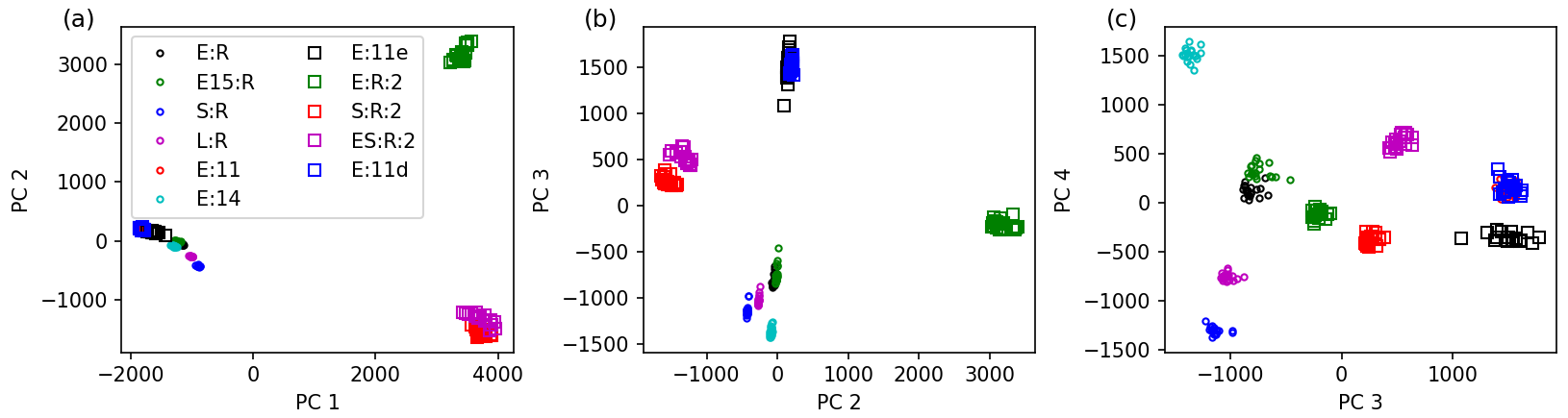}
    \caption{Low-dimensional PC representations of the 2-point statistics of the 11  classes of generated dislocation networks analyzed in Case study 2. (a) PC1-PC2 subspace, (b) PC2-PC3 subspace, and (c) PC3-PC4 subspace.}
    \label{fig:CS2pc}
\end{figure}

Fig. \ref{fig:CS2pc} presents the 2-point statistics of the generated networks in the reduced PC space. As before, in a strictly unsupervised manner, the proposed protocol has successfully identified the inherent structural differences in the ensemble of generated dislocation networks, and has compactly captured these differences in just a few PC scores. Indeed, it is satisfying to observe that PCA has learned to separate each class into its own cluster and that the interclass variance significantly exceeds the intraclass variance. The first 5 PC scores capture roughly 60\% of the total variance in the dataset.  As before, the higher PC scores capture some of the more complex differences responsible for intraclass variance and noise. 

Because of the increased ensemble diversity included in this case study, the interpretation of the features captured by each PC basis is significantly more difficult. This is often the case with the use of PCA for dimensionality reduction in the extraction of material structure statistics \cite{molecdynamics, eutectic}. Even so, the first few PC scores still display interpretable behavior. Fig. \ref{fig:CS2pc}(a) illustrates that the first PC score separates networks with single populated slip systems (on the left) from networks with two populated slip systems (on the right). This is because the values of the auto-correlation at $t=0$ for these two sets of networks exhibit the highest variance, amongst all of the computed spatial statistics for each network for this case study \footnote{Although the total dislocation length was maintained constant, the dislocation density for each slip system has been varied with the inclusion of two slip system networks}. This behavior is consistent with the observations in Case Study 1, and should be generally expected. 

The second PC score appears to capture the differences in the network's dislocation character. Observing Fig. \ref{fig:CS2pc}(b) from left to right, networks containing screw dislocations are present on the left, whereas edge dislocation containing networks are on the right. Again, the protocol has identified that loop containing networks display both behaviors. Interestingly, the two slip system networks are differentiated by the dislocation character on their secondary slip-system. Specifically, ES:R:2 networks, which have edge dislocations on SS1, are categorized on the left side. 

\begin{figure}[h]
    \centering
    \includegraphics[width=84mm]{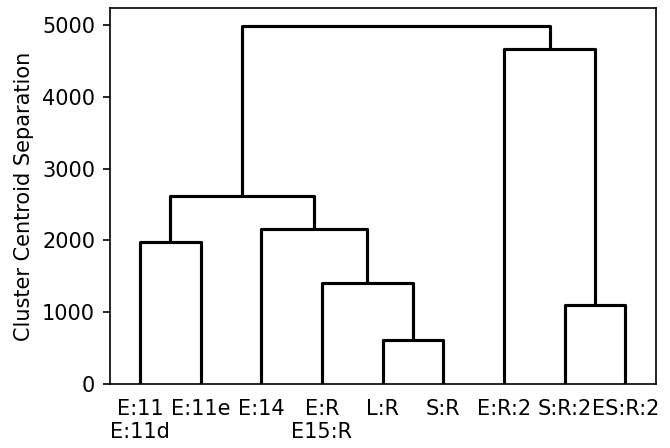}
    \caption{Dendrogram depicting the separation of the 11 classes of dislocation networks based on their first five PC scores.}
    \label{fig:dendroCS2}
\end{figure}

As before, the higher order PC scores become increasingly difficult to interpret. For example, PCs 3 and 4 display limited intuitive structure. As an alternative, the dendrogram is a useful tool for visualizing and interpreting the structure of high dimensional spaces. Fig. \ref{fig:dendroCS2} presents a dendogram showing the distances between the 11 classes of networks based on their first five PC scores. In its construction, the data points are recursively separated into smaller clusters and the inter-cluster distances (defined as euclidean distance between cluster centroids) are depicted. Clusters that are within the same branch, are closer together in the reduced PC space.

Observing the dendrogram, Fig. \ref{fig:dendroCS2}, we can conclude that the reduced PC space displays satisfying and interpretable clustering. The most striking separation, as we might have guessed from our analysis of PC1, is between the class clusters that contain multiple and single populated slip systems. Within each super-cluster, the individual clusters are further ordered. Within the single slip system cluster, the classes containing random global behavior are clustered close together (in the center of the dendrogram). Similarly, those containing uniform global behavior are also closely clustered (on the left side). We note that class E:14, which contains a higher number of uniformly spaced planes is centered, roughly equidistant from these two regimes. Clearly, the dendogram captures the cluster relationships in the five dimensional space much more accurately than what is visualized in the 2-D projections shown in Fig. \ref{fig:CS2pc}. Altogether, these observations indicate that the reduced PC space is well ordered, where the networks that are structurally similar are clustered closer together. 

The classification capability of the PC representations was further verified using the K-Means algorithm \cite{DSP}, which returned 10 separable clusters. The algorithm correctly separated all of the classes except the E:11 and E:11d classes. This is an expected consequence from our choice of the local state descriptor employed in this case study. These two classes are differentiated solely by the sense of individual dislocations, which was not accounted for in the simple scalar descriptor of the local state. This can be remedied by switching to a vectorial local state or, alternatively, adding additional local states describing the dislocation sense. It highlights the importance of selecting adequate local state descriptors for the application. 

\begin{figure}[h]
    \centering
    \includegraphics[width=129mm]{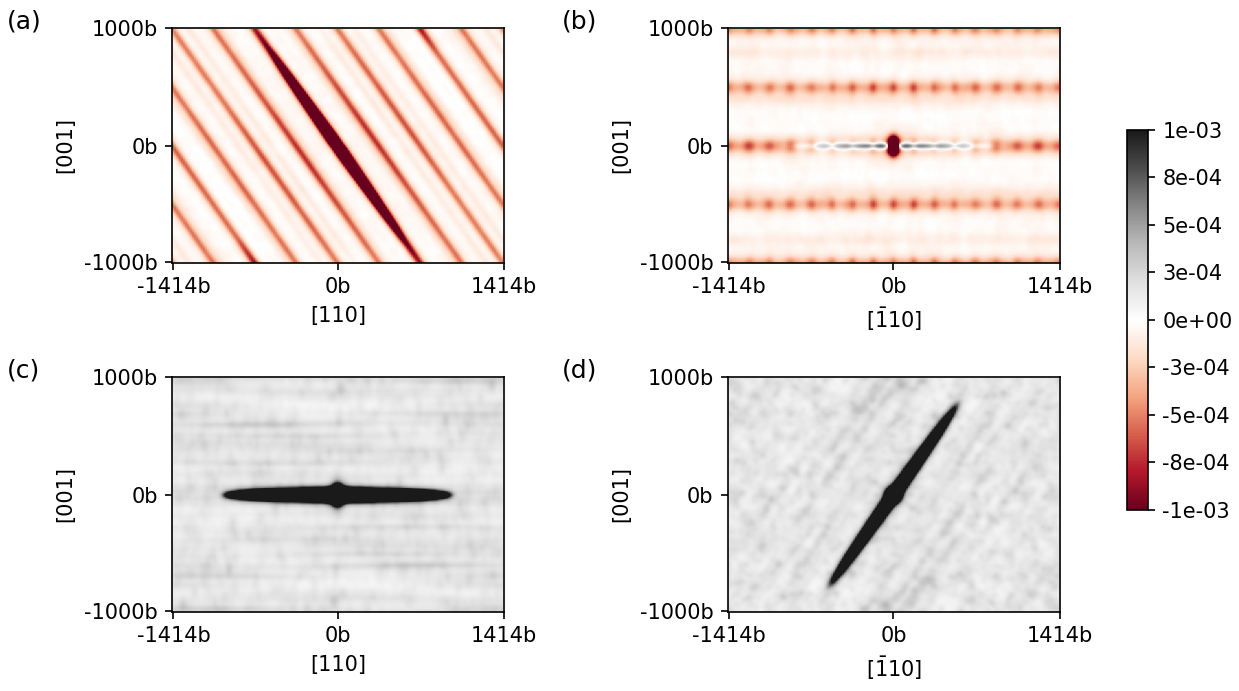}
    \caption{Selected 2-D cross-sections from the auto-correlation components of the first basis vector. (a)  $(\bar{1}10)$ plane of SS1 auto-correlation, (b) $(110)$ plane of SS1 auto-correlation, (c) $(\bar{1}10)$ plane of SS2 auto-correlation, (d) $(110)$ plane of SS2 auto-correlation.}
    \label{fig:basisCS2}
\end{figure}

Analyzing the basis vectors, which contain all the weighted features represented by each PC score, can provide similar insights about the structure of the reduced space. However, they are often challenging to interpret because of their high dimensionality and complex features. Fig. \ref{fig:basisCS2} presents selected 2-D sections of the auto-correlation components of the first PC basis vector for the present case study. Figs. \ref{fig:basisCS2}(a) and \ref{fig:basisCS2}(c) display the $(\bar{1}10)$ cross-sections, while the remaining display the $(110)$ cross-sections.

The PC1 basis vector (see Fig. \ref{fig:basisCS2}) exhibits a large number of high correlation features similar to those discussed earlier. For example, the high intensity features in the centers of these basis maps are clearly capturing certain combinations of the screw and edge characters of dislocations in both the SS1 and SS2 slip systems. Similarly, certain aspects of the global arrangements of the dislocations are also captured in the PC1 basis. For example, Fig. \ref{fig:basisCS2}(a) displays the 11 uniform planes that were populated in the majority of the classes with global structure, while  Fig. \ref{fig:basisCS2}(b) captures the structure of Class E:11e networks displaying uniform spacing within the slip plane. Because of the high values in center of the PC1 basis maps and because none of the previously identified features are specifically targeted, we can conclude that it is capturing a weighted combination of the dislocation densities on both slip systems. Furthermore, we observe that the second autocorrelation map is generally positive, while the first is generally negative. Combining these observations, we can see why the first PC score effectively separates networks by the relative dislocation populations on the first and second slip systems. 

\section{Discussion} \label{discussion}
Although the proposed protocols demonstrated tremendous promise, one needs to pay careful attention to the consequences of the voxelization employed in the computation of the 2-point statistics. In general, correlations calculated on sparse voxelizations (i.e., containing a lot of zeros) will display significant noise because the majority of the voxels will not contribute to the correlation \cite{realdsp, randproc, kalidbook1}. However, because of the 1D nature of individual dislocations, such a voxelization is necessary to adequately separate dislocations and represent their behavior (e.g., shape, curvature). We note that even with these challenges, the significant improvement in computational efficiency offered through the FFT justifies voxelization. 

There exist different strategies to minimize the statistical noise described above. In the proposed protocol, it was partially addressed using kernel smoothing. However, that improvement is largely limited by a need to avoid excessively distorting the shape of individual dislocations. Building on prior  efforts at the mesoscale, \cite{neizgodaSVE,kanit, stochmicro}, two other strategies could be utilized. First, the domain size can be increased to attain a more representative volume. Second, statistics calculated on several independent and identically distributed (i.i.d.) small networks can be averaged for an equivalent effect \cite{neizgodaSVE}. Both strategies have their limitations. In dislocation studies, increasing the domain size comes with significant, insurmountable computational cost \cite{capDDDsurvey}. However, identification of multiple i.i.d. networks can prove equally difficult. This is exacerbated because there are not reliable ways to systematically generate realistic dislocation networks. 

\begin{figure}[h]
    \centering
    \includegraphics[width=174mm]{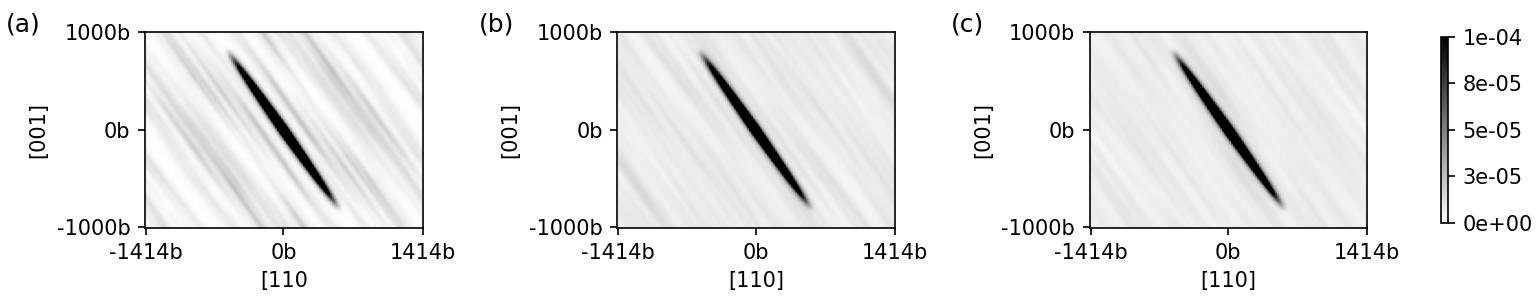}
    \caption{$(110)$ Cross-sections from E:R network SS1 autocorrelation maps. (a) $(2000b)^3$ domain, single network (b) $(4649b)^3$ domain, single network (c) average from twenty $(2000b)^3$ domains.}
    \label{fig:disloc}
\end{figure}

Fig. \ref{fig:disloc} illustrates the benefits of the two solution strategies described above using class E:R networks. For a single E:R network, Fig. \ref{fig:disloc}a, the secondary bands in the auto-correlation maps are erroneously suggesting that the selection of the planes for the placement of the dislocations is not completely random.  These features are significantly reduced when the volume is increased to $(4649b)^3$ at the same overall dislocation density, Fig. \ref{fig:disloc}(b). The artifacts are also removed by the averaging the auto-correlations of twenty i.i.d. networks of the original size. Note that the $t=0$ auto-correlation value for all three plots is the same, indicating that the dislocation density is the same. 

\section{Conclusions}
In this paper, a computationally efficient, flexible framework was presented for calculating salient, low dimensional, spatial statistics for dislocation networks. More specifically, it was shown that Siddon's algorithm and the FFT can be used to calculate efficiently 2-point statistics for discrete dislocation networks. To address the unwieldy dimensionality of the calculated 2-point statistics, it was demonstrated that PCA can be used to extract low-dimensional salient features from the 2-point statistics of an ensemble of networks. To illustrate their application within the data scientific study of dislocation networks, the PC scores were used to differentiate dislocation networks based on their bulk dislocation density as well the local and global behavior of their constituent dislocations. Finally, strategies were discussed to mitigate the statistical noise in the computed  2-point statistics. 

\section{Acknowledgement}
The authors would like to acknowledge funding from ONR N00014-18-1-2879.

\section{Conflict of Interest}
The authors declare that they have no conflict of interest.


\begin{thebibliography}{10}
\providecommand{\url}[1]{{#1}}
\providecommand{\urlprefix}{URL }
\expandafter\ifx\csname urlstyle\endcsname\relax
  \providecommand{\doi}[1]{DOI \discretionary{}{}{}#1}\else
  \providecommand{\doi}{DOI \discretionary{}{}{}\begingroup
  \urlstyle{rm}\Url}\fi

\bibitem{mcdowellplasticity}
D.~McDowell, Int. J. Plast. \textbf{26}, 1280 (2010)

\bibitem{elliot}
J.~Elliot, Int. Mater. Rev. \textbf{56}, 207 (2011)

\bibitem{cpfem}
F.~Roters, P.~Eisenlohr, L.~Hantcherli, D.~Tjahjanto, T.~Bieler, D.~Raabe, Acta
  Mater. \textbf{58}, 1152 (2010)

\bibitem{druckermultiscale}
D.~Drucker, J. Eng. Mater. Technol. \textbf{106}, 286 (1984)

\bibitem{mcdowellviscoplast}
D.~McDowell, Mater. Sci. Eng., R \textbf{62}, 67 (2008)

\bibitem{capDDDsurvey}
R.~LeSar, L.~Capolungo, \emph{In Handbook for Materials Modeling} (Springer,
  Switzerland, 2020)

\bibitem{azabpo}
A.~El-Azab, G.~Po, \emph{In Handbook of Materials Modeling} (Springer,
  Switzerland, 2020)

\bibitem{greer}
J.~Greer, W.~Nix, Phys. Rev. B \textbf{73}, 245410 (2010)

\bibitem{phasefield}
I.~Steinbach, Modell. Simul. Mater. Sci. Eng. \textbf{17}, 073001 (2009)

\bibitem{PCevolution}
Y.~Yabansu, P.~Steinmetz, J.~Hotzer, S.~Kalidindi, B.~Nestler, Acta Mater.
  \textbf{124}, 182 (2017)

\bibitem{kalidbook1}
B.~Adams, S.~Kalidindi, D.~Fullwood, \emph{Microstructure Sensitive Design for
  Performance Optimization} (Butterworth-Heinemann, Waltham, MA, 2013)

\bibitem{fullwoodsurvey}
D.~Fullwood, S.~Niezgoda, B.~Adams, S.~Kalidindi, Prog. Mater. Sci.
  \textbf{55}, 477 (2010)

\bibitem{pymks}
D.~Brough, D.~Wheeler, S.~Kalidindi, Integr. Mater. Manuf. Innov. \textbf{6},
  36 (2017)

\bibitem{tem}
G.~Liu, S.~House, J.~Kacher, M.~Tanaka, K.~Higashida, I.~Robertson, Mater.
  Charact. \textbf{87}, 1 (2014)

\bibitem{kachertem}
T.~Ruggles, Y.~Yoo, B.~Dunlap, M.~Crimp, J.~Kacher, Ultramicroscopy
  \textbf{210}, 112927 (2020)

\bibitem{originaltem}
J.~Barnard, J.~Sharp, J.~Tong, P.~Midgley, Science \textbf{313}, 319 (2006)

\bibitem{torquato}
S.~Torquato, \emph{Random Heterogeneous Materials} (Springer, New York, NY,
  2002)

\bibitem{hirthlothe}
J.~Hirth, J.~Lothe, \emph{Theory of Dislocations} (Krieger, Malabar, FA, 1982)

\bibitem{hullbacon}
D.~Hull, D.~Bacon, \emph{Introduction to Dislocations} (Butterworth-Heinemann,
  Burlington, MA, 2011)

\bibitem{taylorrule}
P.~Franciosi, M.~Berveiller, A.~Zaoui, Acta Metall. \textbf{28}, 273 (1980)

\bibitem{zaiser}
M.~Zaiser, K.~Bay, P.~Hahner, Acta Mater. \textbf{47}, 2463 (1999)

\bibitem{AzabDeng}
J.~Deng, A.~El-Azab, Modell. Simul. Mater. Sci. Eng. \textbf{17}, 075010 (2009)

\bibitem{wang}
H.~Wang, R.~Lesar, J.~Rickman, Philos. Mag. A \textbf{78}, 1195 (1998)

\bibitem{kacher}
C.~Landon, B.~Adams, J.~Kacher, J. Eng. Mater. Technol. \textbf{130}, 021004
  (2008)

\bibitem{anderson}
J.~Anderson, A.~El-Azab (2020)

\bibitem{groma1}
I.~Groma, P.~Balogh, Acta Mater. \textbf{47}, 3647 (1999)

\bibitem{groma2}
I.~Groma, F.~Csikor, M.~Zaiser, Acta Mater. \textbf{51}, 1271 (2003)

\bibitem{orig2PS}
D.~Fullwood, S.~Kalidindi, B.~Adams, \emph{Chapter 13 in Electron Backscatter
  Diffraction in Materials Science} (Springer, Boston, MA, 2009)

\bibitem{taxonomy}
T.~Fast, O.~Wodo, B.~Ganapathysubramanian, S.~Kalidindi, Acta Mater.
  \textbf{108}, 176 (2016)

\bibitem{eutectic}
A.~Choudhury, Y.~Yabansu, S.~Kalidindi, A.~Dennstedt, Acta Mater. \textbf{110},
  131 (2016)

\bibitem{molecdynamics}
S.~Kalidindi, J.~Gomberg, Z.~Trautt, C.~Becker, Nanotechnology \textbf{26}
  (2015)

\bibitem{nonlinear2ps}
M.~Latypov, L.~Toth, S.~Kalidindi, Comput. Methods Appl. Mech. Engrg.
  \textbf{346}, 180 (2019)

\bibitem{fatigue2ps}
N.~Paulson, M.~Priddy, D.~McDowell, S.~Kalidindi, Int. J. Fatigue \textbf{119},
  1 (2019)

\bibitem{polycrystal2ps}
N.~Paulson, M.~Priddy, D.~McDowell, S.~Kalidindi, Acta Mater. \textbf{129}, 428
  (2017)

\bibitem{matthew}
M.~Barry, K.~Wise, S.~Kalidindi, S.~Kumar, J. Phys. Chem. Lett. \textbf{11},
  9093 (2020)

\bibitem{kroner}
E.~Kroner, \emph{Statistical Continuum Mechanics} (Springer, New York, NY,
  1972)

\bibitem{azabStochFibProc2006}
A.~El-Azab, Scr. Mater. \textbf{54}, 723 (2006)

\bibitem{stochmicro}
S.~Niezgoda, Y.~Yabansu, S.~Kalidindi, Acta Mater. \textbf{59}, 6387 (2011)

\bibitem{randproc}
A.~Papoulis, S.~Pillai, \emph{Probability, Random Variables, and Stochastic
  Processes} (McGraw Hill, Chennai, 2002)

\bibitem{sauzay}
M.~Sauzay, L.~Kubin, Prog. Mater. Sci. \textbf{56}, 725 (2011)

\bibitem{temporal_stat}
J.~Deng, A.~El-Azab, Philos. Mag. \textbf{90}, 3651 (2010)

\bibitem{hochrainer}
T.~Hochrainer, M.~Zaiser, P.~Gumbsch, Philos. Mag. \textbf{87}, 1261 (2007)

\bibitem{sedlacek}
R.~Sedlacek, C.~Schwarz, J.~Kratochvil, E.~Werner, Philos. Mag. \textbf{87},
  1225 (2007)

\bibitem{giessen}
S.~Limkumnerd, E.V. der Giessen, Phys. Rev. B \textbf{77}, 184111 (2008)

\bibitem{valdenaire}
P.~Valdenaire, Y.L. Bouar, B.~Appolaire, A.~Finel, Phys. Rev. B \textbf{93},
  214111 (2016)

\bibitem{arsenlis}
A.~Arsenlis, D.~Parks, Acta Mater. \textbf{47}, 1597 (1999)

\bibitem{kroner_dislocation}
E.~Kroner, Int. J. Solids Struct. \textbf{38}, 1115 (2001)

\bibitem{AzabXia2015}
S.~Xia, A.~El-Azab, Modell. Simul. Mater. Sci. Eng. \textbf{23}, 055009 (2015)

\bibitem{bertin}
N.~Bertin, Int. J. Plast. \textbf{122}, 268 (2019)

\bibitem{siddons}
G.~Han, Z.~Liang, J.~You, IEEE Nucl. Sci. Symp. Conf. Rec. p. 1515 (2000)

\bibitem{princeton}
J.~Shlens.
\newblock A tutorial of principal component analysis.
\newblock
  \urlprefix\url{https://www.cs.princeton.edu/picasso/mats/PCA-Tutorial-Intuition\_jp.pdf}.
\newblock Accessed 28 November 2020.

\bibitem{DSP}
T.~Hastie, R.~Tibshirani, J.~Friedman, \emph{The Elements of Statistical
  Learning} (Springer, New York, NY, 2016)

\bibitem{Annihil}
U.~Essmann, H.~Mughrabi, Philos. Mag. A \textbf{40}, 731 (1979)

\bibitem{ahmet}
A.~Cecen, \emph{Calculation, utilization, and inference of spatial statistics
  in practical spatio-temporal data} (Georgia Tech Library, Atlanta, GA, 2017)

\bibitem{imagesmooth}
A.~Makandar, B.~Halalli, Int. J. Comput. Appl. p. 109 (2015)

\bibitem{alsmooth}
A.~Iskakov, S.~Kalidindi, Integr. Mater. Manuf. Innov. \textbf{9}, 70 (2020)

\bibitem{cai}
W.~Cai, A.~Arsenlis, C.~Weinberger, V.~Bulatov, J. Mech. Phys. Solids
  \textbf{54}, 561 (2006)

\bibitem{neizgoda}
S.~Neizgoda, D.~Fullwood, S.~Kalidindi, Acta Mater. \textbf{56}, 5285 (2008)

\bibitem{ekg}
M.~Chawla, Appl. Soft Comput. \textbf{11}, 2216 (2011)

\bibitem{realdsp}
M.~Vetterli, J.~Kovacevic, V.~Goyal, \emph{Foundations of Signal Processing}
  (Cambridge University Press, Cambridge, UK, 2014)

\bibitem{neizgodaSVE}
S.~Niezgoda, D.~Turner, D.~Fullwood, S.~Kalidindi, Acta Mater. \textbf{58},
  4432 (2010)

\bibitem{kanit}
T.~Kanit, S.~Forest, I.~Galliet, V.~Mounoury, D.~Jeulin, Int. J. Solids Struct.
  \textbf{40}, 3647 (2003)

\end{thebibliography}
\end{document}